# Tunable optical bistability in grapheme Tamm plasmon/Bragg reflector hybrid structure at terahertz frequencies

Shenping Wang[1], Jiao Xu[1], Hongxia Yuan[1], Huayue Zhang[1], Xin Long[1], Leyong Jiang[1, *], and Jie Jiang[2]

[1]School of Physics and Electronics, Hunan Normal University, Changsha 410081, China;

[2]Hunan Key Laboratory of Super Microstructure and Ultrafast Process, School of Physics and Electronics, Central South University, Changsha 410083, China；

Corresponding Author: * jiangly28@hunnu.edu.cn

**Abstract**：We propose a composite multilayer structure consist of graphene Tamm plasmon and Bragg reflector with defect layer to realize the low threshold and tunable optical bistability (OB) at the terahertz frequencies. This low-threshold OB originates from the couple of the Tamm plasmon (TP) and the defect mode (DM). We discuss the influence of graphene and the DM on the hysteretic response of the TM-polarized and TE-polarized reflected light. It is found that the switch-up and switch-down threshold required to observe the optical bistable behavior are lowered markedly due to the excitation of the TP and DM. Besides, the switching threshold value can be further reduced by coupling the TP and DM. We believe these results will provide a new avenue for realizing the low threshold and tunable optical bistable devices and other nonlinear optical devices.

# 1. Introduction

Optical bistability refers to the nonlinear optical effect that corresponds to two stable output light intensity states under a given input light intensity, and the two states can be converted to each other through optical control [1].Optical bistability has potential and has wide applications in micro nan optoelectronic devices such as all-optical logic gate [2], all-optical switching [3], optical memory [4] and optical diode [5] because of its main characteristics such as hysteresis and mutation . At present, the requirements of optical communication systems for the switching speed of all-optical bistable devices and optical switches are not too high, so reducing the threshold power has become the main goal of the research of optical bistable and all-optical switches . At the moment, the reduction of threshold is mainly carried out in two ways: the first way is to design micro nano structures with local field enhancement effect to realize the local enhancement of electric field and create conditions for the reduction of threshold, such as microring resonator [6], slit waveguide [7], nonlinear photonic crystals [8,9] Metamaterials [10], surface plasmons [11-13], metal dielectric multilayers [14], metal gratings [15], *etc*. The second method is to find and combine optical materials with high nonlinear coefficients. It is worth mentioning that graphene, the representative of two-dimensional materials emerging in recent years, has received extensive attention in the study of optical bistability due to its large nonlinear refractive index and conductivity, such as one-dimensional (1D) periodic structure [16], two-dimensional photonic crystal cavity [17], multilayer dielectric Bragg mirror [18] or plasma nanostructure [19], These structures provide a way for the practical application of optical bistable devices.

  Tamm plasmin (TPs) [20] is a non-dissipative local interface mode. TPs will enhance the in-situ at the boundary of different materials, and the field localization intensity will gradually weaken with the distance away from the interface region [21].Compared with the surface plasmon [22], TPs can be polarized by both TM and TE, and it does not need to meet the specific incident angle [23]. In addition, it also has strong localization, so it is widely used in many fields. In addition, the traditional TPs is mainly excited based on the metal Bragg mirror structure [24]. Because graphene has metal like properties under some conditions [25], in recent years, TPs Based on the graphene Bragg mirror structure and its low threshold bistability have also attracted attention [26]. In these works, graphene supports excitation of TPs to realize local field enhancement and provide large nonlinear materials for the whole structure. Therefore, it has obvious advantages in realizing with a low threshold. However, for some common TPs layered structures, there are often more than a single TPs mode under specific conditions. If these modes are coupled, will there be optical bistability? This is also a very interesting question. Based on this, in this paper, we study the coupling of Tamm plasmin and defect modes and the corresponding optical bistability in graphene-covered photonic crystal multilayer structures. We find that by controlling the coupling of Tamm plasmin and defect mode in terahertz band, we can not only achieve tunable optical bistability with a lower threshold, but also be accompanied by some interesting phenomena. The effects of graphene and defect modes on hysteresis curves under TM polarization are discussed. The results show that the introduction of the defect mode has a positive effect on the reduction of threshold. At the same time, the large third-order nonlinear conductivity of graphene not only provides a key premise for the generation of bistable curve, but also its dynamically controllable linear conductivity provides a feasible way for the dynamic regulation of bistable curves. In addition, the Fano resonance generated by mode coupling makes the whole structure support both transmission and reflection bistable curves. The numerical simulation further supports the theoretical results. We believe that the mode coupling bistable scheme based on graphene provides a novel scheme for the realization of tunable optical bistable devices and other nonlinear optical devices.

## 2. Theoretical Model and Method

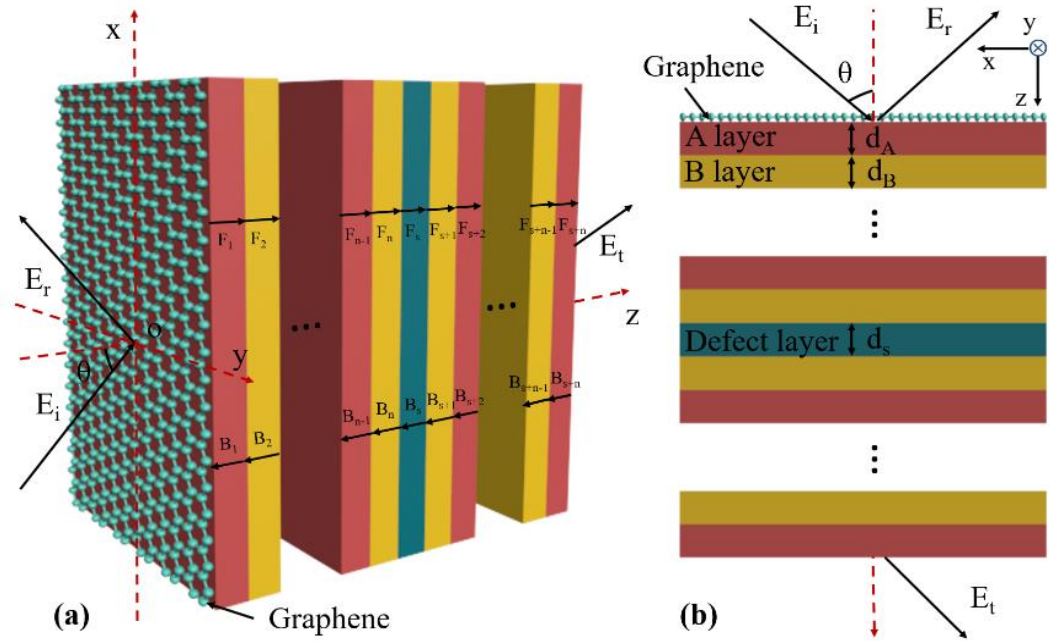

Fig. 1. Schematic diagram of photonic crystal multilayer structure with graphene and defect layer.

We consider a composite structure composed of graphene and one-dimensional photonic crystals. A defect layer is added between two symmetrical one-dimensional photonic crystals. At the same time, one end of the overall structure is covered with a layer of graphene, as shown in Fig. 1. The photonic crystals on both sides of the defect layer are formed by alternating superposition of dielectric layers A and B with period N.

The plane wave with amplitude $E_i$ is incident on the sandwich structure at the incident angle $\theta$ to produce reflected and transmitted waves with amplitudes $E_r$ and $E_t$ respectively. $\varepsilon_A$, $\varepsilon_B$ and $\varepsilon_s$ are the dielectric constants of layers A, B and defect respectively, the thickness layer is represented by $d_A$、$d_B$ and $d_s$ respectively, and the grating period number is N. For graphene, because of its monatomic layer characteristics, we use conductivity to express its characteristics. Without considering the applied magnetic field and in the case of random phase approximation, the surface conductivity of graphene in the terahertz band can be approximately expressed as [27]:

$$\sigma_0 \approx \frac{ie^2 E_F}{\pi \hbar^2 (\omega + i/\tau)}, \tag{1}$$

where e, $\omega$, $v_F \approx 10^6 \, m/s$ and $n_{2D}$ represent charge, angular frequency, electron Fermi velocity and carrier density respectively; $E_F$ is the Fermi level of graphene, and $E_F = \hbar v_F \sqrt{\pi n_{2D}}$; In addition, the third-order nonlinear conductivity of graphene can be expressed as [28]:

$$\sigma_3 = -i\frac{9}{8} \frac{e^4 v_F^2}{\pi \hbar^2 E_F \omega^3}, \tag{2}$$

therefore, considering the nonlinearity, the conductivity of graphene can be expressed as the expression related to the electric field at the graphene interface: $\sigma = \sigma_0 + \sigma_3 |E|^2$. It can be seen from the conductivity expression of graphene that the conductivity of graphene has flexible electrically adjustable characteristics due to the close relationship between $E_F$ and carrier density $n_{2D}$. In this paper, we set the center wavelength to $\lambda_0 = 300 \, \mu m$. For photonic crystals, we select glass and dielectric materials, and the initial parameters are set to $d_A = 30.5 \, \mu m$, $d_B = 45.5 \, \mu m$, $N = 20$, $n_A = 2.3$ and $n_B = 1.5$ respectively. The initial parameters of graphene are $E_F = 0.9 \, eV$, $\tau = 1.2 \, ps$. For the defect layer, we select the material Si, and the initial parameters are $d_s = 48.75 \, \mu m$, $n_s = 3.416$.

We choose the z-axis as the propagation direction and set the position of graphene as $z = 0$. At the same time, the x-axis is parallel to the plane of graphene. At this time, the conductivity of graphene can be expressed as $\sigma = \sigma_0 + \sigma_3 |E_{0y}(z=0)|^2$. In this paper, we focus on the bistability of TM polarization. When the incident angle is $\theta$, the electric and magnetic fields in the air on the left of graphene can be expressed as:

$$\begin{cases} H_{0y} = H_i e^{ik_{iz}z} e^{ik_x x} + H_r e^{-ik_{iz}z} e^{ik_x x}, \\ E_{0x} = \dfrac{k_{iz}}{\varepsilon_i \varepsilon_0 \omega} H_i e^{ik_{iz}z} e^{ik_x x} - \dfrac{k_{iz}}{\varepsilon_i \varepsilon_0 \omega} H_r e^{-ik_{iz}z} e^{ik_x x}, \\ E_{0z} = -\dfrac{k_x}{\varepsilon_i \varepsilon_0 \omega} H_i e^{ik_{iz}z} e^{ik_x x} - \dfrac{k_x}{\varepsilon_i \varepsilon_0 \omega} H_r e^{-ik_{iz}z} e^{ik_x x}, \end{cases} \quad (3)$$

where $k_0 = \omega/c$, $k_{0z} = k_0 \cos(\theta)$, $k_x = k_0 \sin(\theta)$ and $\mu_0$ are the permeability of free space; $E_i$ and $E_r$ are the amplitudes of incident electric field and reflected electric field respectively. Similarly, the electric and magnetic fields of the defect layer can be expressed as:

$$\begin{cases} H_{25y} = F_{25} e^{ik_{sz}(z-(12d_A+12d_B))} e^{ik_x x} + B_{25} e^{-ik_{sz}(z-(12d_A+12d_B))} e^{ik_x x}, \\ E_{25x} = \dfrac{k_{sz}}{\varepsilon_s \varepsilon_0 \omega} F_{25} e^{ik_{sz}(z-(12d_A+12d_B))} e^{ik_x x} - \dfrac{k_{sz}}{\varepsilon_s \varepsilon_0 \omega} B_{25} e^{-ik_{sz}(z-(12d_A+12d_B))} e^{ik_x x}, \\ E_{25z} = -\dfrac{k_x}{\varepsilon_s \varepsilon_0 \omega} F_{25} e^{ik_{sz}(z-(12d_A+12d_B))} e^{ik_x x} - \dfrac{k_x}{\varepsilon_s \varepsilon_0 \omega} B_{25} e^{-ik_{sz}(z-(12d_A+12d_B))} e^{ik_x x}. \end{cases} \quad (4)$$

In medium m ($m = \{2,3,...n\}$), the electric field and magnetic field can be expressed as:

$$\begin{cases} H_{my} = F_m e^{ik_{\varsigma z}(z-d)} e^{ik_x x} + B_m e^{-ik_{\varsigma z}(z-d)} e^{ik_x x}, \\ E_{mx} = \dfrac{k_{\varsigma z}}{\varepsilon_\varsigma \varepsilon_0 \omega} F_m e^{ik_{\varsigma z}(z-d)} e^{ik_x x} - \dfrac{k_{\varsigma z}}{\varepsilon_\varsigma \varepsilon_0 \omega} B_m e^{-ik_{\varsigma z}(z-d)} e^{ik_x x}, \\ E_{mz} = -\dfrac{k_x}{\varepsilon_\varsigma \varepsilon_0 \omega} F_m e^{ik_{\varsigma z}(z-d)} e^{ik_x x} - \dfrac{k_x}{\varepsilon_\varsigma \varepsilon_0 \omega} B_m e^{-ik_{\varsigma z}(z-d)} e^{ik_x x}, \end{cases} \quad (5)$$

in the above formula, when m is an odd number $\varsigma = A$, when m is an even number $\varsigma = B$, d represents the distance from the dielectric layer to $z = 0$, $k_{jz} = \sqrt{k_0^2 n_j^2 - k_x^2}$, $j = \{s, A, B\}$. On the rightmost side of the whole structure, the electric field and magnetic field can be expressed as:

$$\begin{cases} H_{(n+1)y} = H_T e^{ik_{Az}(z-(d_s + N(d_A + d_B)))} e^{ik_x x}, \\ E_{(n+1)x} = \dfrac{k_{Az}}{\varepsilon_A \varepsilon_0 \omega} H_T e^{ik_{Az}(z-(d_s + N(d_A + d_B)))} e^{ik_x x}, \\ E_{(n+1)z} = -\dfrac{k_x}{\varepsilon_A \varepsilon_0 \omega} H_T e^{ik_{Az}(z-(d_s + N(d_A + d_B)))} e^{ik_x x}, \end{cases} \quad (6)$$

where $H_t$ is the amplitude of the transmitted electric field.

For boundary conditions, the following conditions are satisfied at $z = d$: $E_{my}(z=d) = E_{(m+1)y}(z=d)$ and $H_{mx}(z=d) = H_{(m+1)x}(z=d)$. In particular, at $z = 0$: $E_{0y}(z=0) = E_{1y}(z=0)$ and $H_{0x}(z=0) - H_{1x}(z=0) = \sigma E_{0y}(z=0)$. Based on the boundary conditions of different contact surfaces, we can get the relationship between $E_i$, $E_r$, $E_t$, $F_m$, $B_m$ ($m = \{2, 3, ...n\}$), so we can get the relationship between $E_r$ and $E_i$, so OB can be observed under appropriate parameter conditions.

## 3. Results and Discussions

When the plane wave is vertically incident on the defective photonic crystal in the horizontal direction, we use the transfer matrix method to correlate the electromagnetic fields on each end face of the dielectric through the transfer matrix, so as to obtain the relationship between reflectivity and frequency.

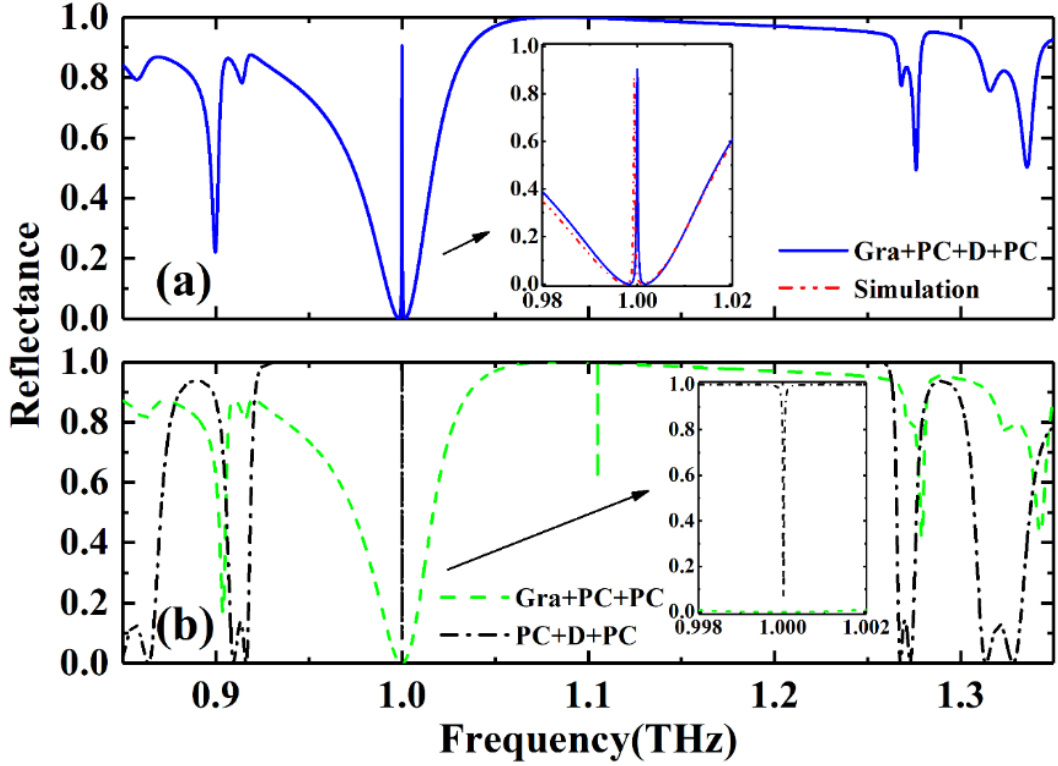

Fig. 2. (a) The double dotted line and solid line are the simulation and matlab calculation results respectively, (b) the function of reflectivity without graphene or defect layer with respect to frequency, and other parameters are set as $N=24$, $d_A=30.5\,\mu m$, $d_B=40.5\,\mu m$, $d_s=48.75\,\mu m$, $\theta=0°$.

We know that TPs can be excited by both TE polarization and TM polarization. In order to simplify the discussion, in the following work, we mainly discuss the TM polarization and set the initial incident angle to $\theta=0°$. In Fig. 2(b), it can be seen that when there is no defect layer in the structure, there is an obvious depression in the reflection spectrum at the frequency of 1THz. When the thickness of the defect layer is set to $d_s=48.75\,\mu m$, the defect mode in the middle of the grating can be excited at the reflection peak ($f=1\,THz$), and this spectral response can be regarded as a typical impedance-like effect.

In order to verify the theoretical results, we use COMSOL software to simulate the optical response in a multilayer system. As shown in Fig. 2(a), the simulation results are very consistent with the theoretical calculation results. When there is no graphene in the structure, because the whole structure cannot excite TPs, there is an obvious photonic band gap in the wavelength range near the incident wavelength.

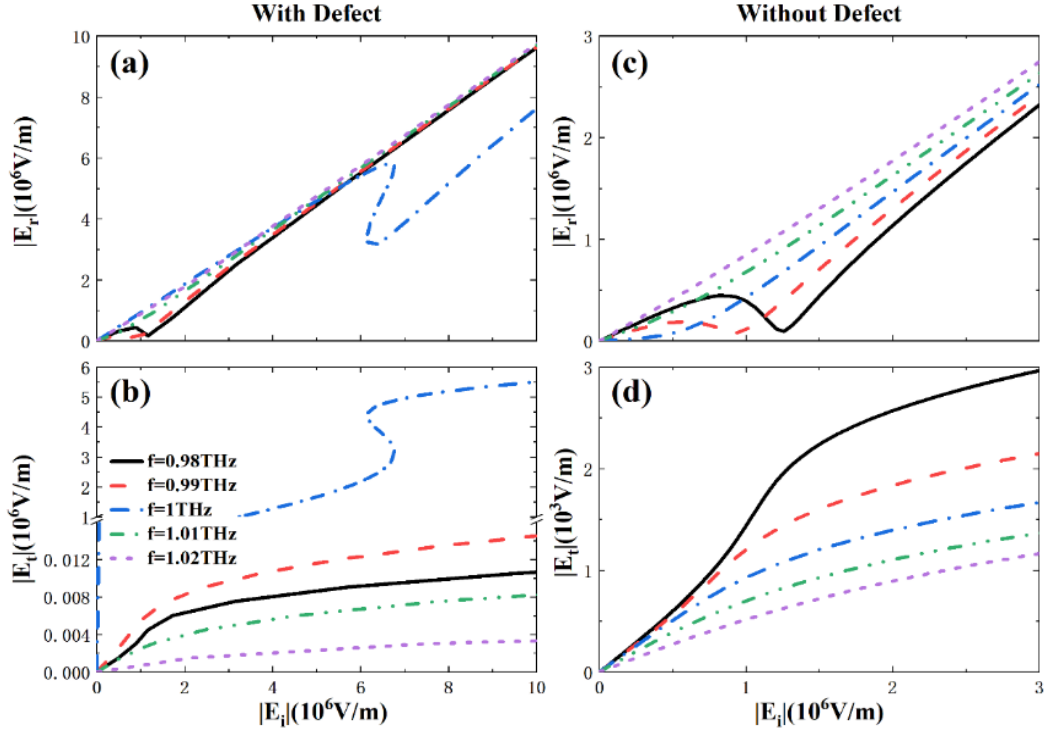

Fig. 3. Between 0.98-1.02 THz, bistable with defect layer (a) reflection type (b) transmission type and without defect layer (c) reflection type (d) transmission type

Next, we discuss the change of behavior of OB around 1THz. Based on the calculation method in the previous part, we show the relationship between reflected electric field $E_r$, transmitted electric field $E_t$ and incident electric field $E_i$, as shown in Fig. 3. We select $\lambda = 300\,\mu m$, and other parameters are the same as those in Figure 2. According to the conductivity formula of graphene in the previous chapter, it can be known that the nonlinearity of graphene decreases with the increase of frequency, which will also gradually narrow the hysteresis width and eventually lead to the disappearance of bistable phenomenon. We can find that the bistability phenomenon from 0.98 THz to 1.02 THz has disappeared, and the trend of bistability is less and less obvious. However, when we add a defect layer to the structure, bistability can occur at 1THz, such as the reflective bistability in Fig. 3(a) and the transmission bistability in Fig. 3(b), which also verifies the function of the reflectivity of the defect layer in Fig. 2(a) as a function of frequency. The third-order nonlinearity of graphene plays a key role in the generation of bistability. At the same time, the photonic crystal multilayer structure with graphene and defect layer plays an active role in better regulating bistability by exciting TPs.

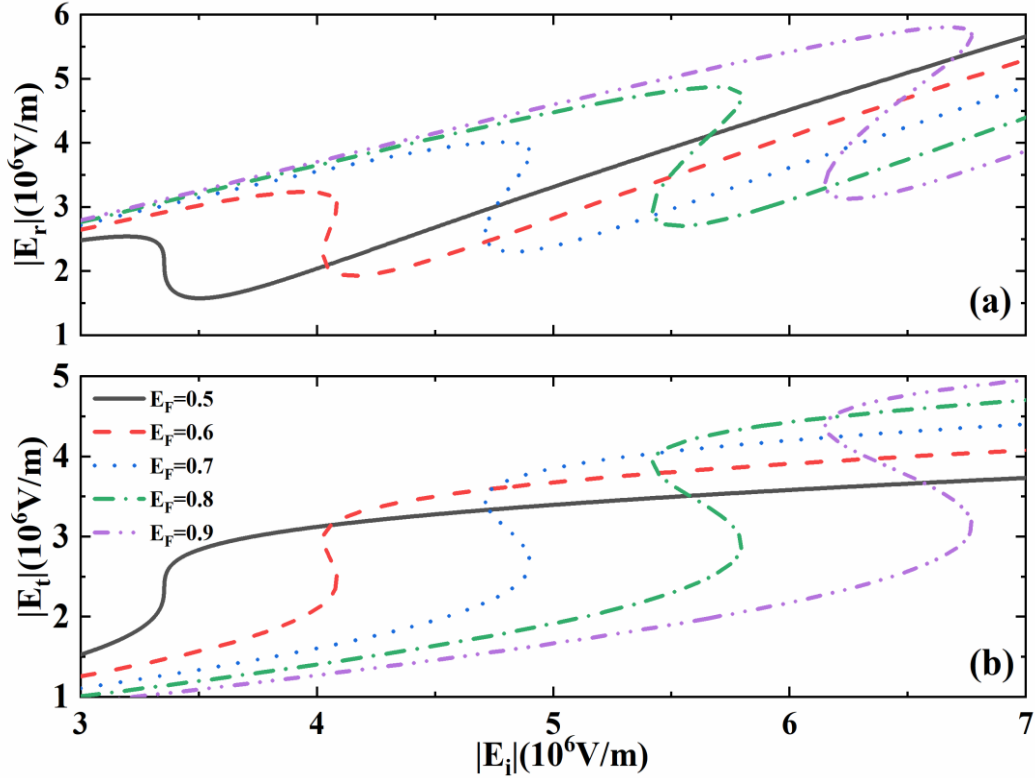

Fig. 4. (a) Dependence of reflected electric field and (b) transmitted electric field on incident electric field at different Fermi levels

In order to study the optical bistability in this structure, when $f = 1\,\text{THz}$ selects different graphene Fermi levels $E_F = 0.5\,\text{eV}、0.6\,\text{eV}、0.7\,\text{eV}、0.8\,\text{eV}、0.9\,\text{eV}$, as shown in Fig. 4, other parameters are the same as those in Fig. 2. In Fig. 4(a), we take $E_F = 0.9\,\text{eV}$ as an example. When the incident electric field $E_i$ is at a small value, the reflected electric field $E_r$ increases slowly with the increase of $E_i$, which is one of the steady states.

When the incident electric field continues to increase and the reflected electric field reaches $E_r = 6.75 \times 10^6 \, V/m$, it will jump to another steady state. If the incident electric field is reduced at this time, the reflected electric field will not immediately return to the first steady state, but will slowly decrease with the incident electric field, and will not jump from the second stable state to the first stable state until the reflected electric field is reduced to $E_r = 6.15 \times 10^6 \, V/m$. Similarly, it can explain the transmission electric field. In addition, the excellent electrical conductivity of graphene also makes it possible to realize tunable optical bistability based on graphene-excited TPs. In Fig. 4, the increase of the Fermi level of graphene makes the hysteresis curve move to a lower incident electric field as a whole, and the hysteresi width begins to shrink and narrow at the same time. It can be seen that although the reduction of Fermi level plays a positive role in further reducing the threshold of optical bistability, it also narrows or even disappears the hysteresis width.

In this part, we first discuss the influence of the presence or absence of graphene on the behavior of OB in Fig. 5(a). In the calculation, we select $f = 1 \, THz$ transmission bistability as an example. From Fig. 5(a), it can be found that the absence of graphene in the structure creates a simple monotonic increasing relationship between transmission electric field $E_t$ and incident electric field $E_i$, That is, the hysteresis curve cannot appear. After graphene is loaded, the hysteresis curve of transmission electric field $E_t$ and incident electric field $E_i$ can be clearly generated, mainly because the huge third-order nonlinear conductivity of graphene itself satisfies the multivalued phenomenon between transmission electric field $E_t$ and incident electric field $E_i$. Next, the relationship between the incident electric field and the transmitted electric field under different graphene layers under certain other conditions is discussed, as shown in Fig. 5(b).

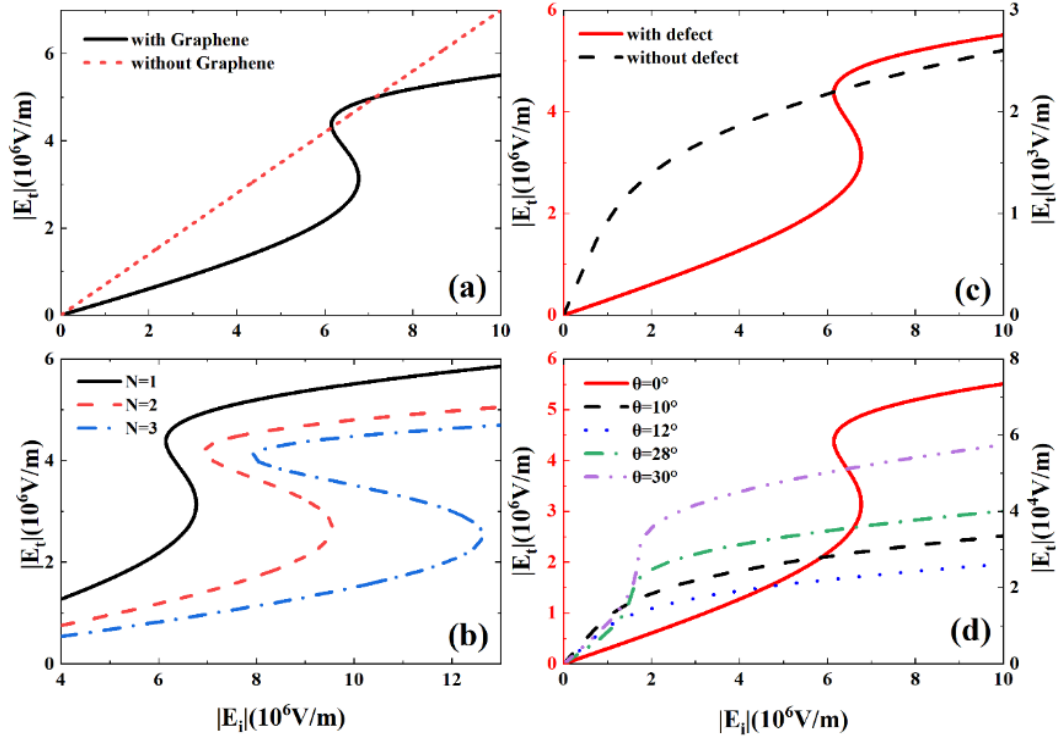

Fig. 5. (a) The dependence of the transmitted electric field on the incident electric field with or without graphene. (b) Dependence of transmitted electric field on incident electric field under different graphene film layers. (c) The dependence of transmitted electric field on incident electric field with or without defect layer. (d) Dependence of transmitted electric field on incident electric field at different incident angles

At $N=1、2、3$, the on thresholds are $2.2\times10^6$ V/m, $2.8\times10^6$ V/m and $3\times10^6$ V/m respectively, and the off thresholds are $1.77\times10^6$ V/m, $2\times10^6$ V/m and $2.15\times10^6$ V/m respectively. From this data, we can see that the on threshold is greater than the off threshold. With the increase of the number of graphene layers, the switching threshold gradually increases. It can also be seen that the range between switching thresholds increases and the optical bistable working range increases. In contrast, after the introduction of graphene, the optical bistability of this structure is more sensitive to the number of layers of graphene. On the premise of the same parameters, we discuss the influence of the defect layer on the behavior of OB. In Fig. 5 (c), we can see that although the nonlinear material graphene can produce the trend of optical bistability, it can produce optical bistability more easily with the addition of defect layer. Finally, shift the line of sight from the structural parameters to the parameters of the incident light, and continue to explore the relationship between the incident electric field and the transmitted electric field under different incident angles with certain other parameters. The effect of incident angle on hysteresis behavior is reflected in both threshold and hysteresis width. In contrast, after the introduction of graphene, the optical bistability of this structure is still very sensitive to the angle of incident light. For the transmitted electric field, with the increase in incident angle, the trend from bistability to bistability is less and less obvious in $0°-12°$. When the angle is $12°-30°$, the bistability trend is more and more obvious, but there is no obvious bistability. In general, compared with traditional metals, graphene-excited TPs can flexibly control the threshold and hysteresis width, which provides a practical idea for the realization of dynamically controllable optical bistable devices.

## 4. Conclusion

In conclusion, we study the optical bistability of light reflected on graphene distributed Bragg structure. By using the coupling of Tamm plasmon and defect mode in graphene covered photonic crystal multilayer structure, we can realize tunable optical bistable devices with a low threshold in terahertz band. The effects of graphene and defect mode on hysteresis curve under TM polarization are discussed. The results show that the introduction of the defect mode has a positive effect on the reduction of threshold. The optical bistability threshold can also be further controlled by the incident angle, the number of graphene layers and Fermi energy. We believe that the graphene optical bistable pathway studied in this paper has broad application prospects, and has potential application prospects in nonlinear optical devices such as optical logic and all-optical switches.

**Acknowledgments**

This work was supported by the National Natural Science Foundation of China (Grant

No. 11704119), the Hunan Provincial Natural Science Foundation of China (Grant No. 2018JJ3325), Scientific Research Fund of Hunan Provincial Education Department (Grant No. 21B0048) and National College Students' innovation and entrepreneurship training program (Grant No. 202110542014).